\documentclass[12pt,fleqn,a4paper]{article}

\usepackage{graphicx}
\usepackage{amsmath,amssymb}

\usepackage[english]{babel}

\usepackage{mcite}

\newcommand{\ttbar}{t\overline{t}}
\newcommand{\qqbar}{q\overline{q}}
\newcommand{\ppbar}{p\overline{p}}
\newcommand{\pt}{p_{\rm T}}
\newcommand{\Oa}{{\cal O}(\alpha)}

\newcommand{\Oas}{{\cal O}(\alpha^{\mathrm 2})}

\newcommand{\Oaas}{{\cal O}(\alpha\alpha_{\mathrm s})}
\newcommand{\Oass}{{\cal O}(\alpha_{\mathrm s}^{\mathrm 2})}
\newcommand{\Oasc}{{\cal O}(\alpha_{\mathrm s}^{\mathrm 3})}

\newcommand{\Oaass}{{\cal O}(\alpha\alpha_{\mathrm s}^{\mathrm 2})}

\newcommand{\loglambda}{\ln\left(\frac{2 \Delta E}{\lambda}\right)}
\newcommand{\logbeta}{\ln\left(\frac{1-\beta}{1+\beta}\right)}
\newcommand{\logbetas}{\ln^2\left(\frac{1-\beta}{1+\beta}\right)}
\newcommand{\logmq}{\ln\left(\frac{m_q^2}{\hat s}\right)}
\newcommand{\logmqs}{\ln^2\left(\frac{m_q^2}{\hat s}\right)}
\newcommand{\li}{{\rm Li}_2\left(\frac{2\beta}{1+\beta}\right)}
\newcommand{\logtu}{\ln\left(\frac{m_t^2-\hat u}{m_t^2-\hat t}\right)}
\newcommand{\litplus}{{\rm Li}_2\left(1-\frac{\hat s(1+\beta)}{2(m_t^2-\hat t)
}\right)}
\newcommand{\litmin}{{\rm Li}_2\left(1-\frac{\hat s(1-\beta)}{2(m_t^2-\hat t)
}\right)}
\newcommand{\liuplus}{{\rm Li}_2\left(1-\frac{\hat s(1+\beta)}{2(m_t^2-\hat u)
}\right)}
\newcommand{\liumin}{{\rm Li}_2\left(1-\frac{\hat s(1-\beta)}{2(m_t^2-\hat u)
}\right)}
\newcommand{\sq}{\sqrt{\hat{s}} }

\begin{document}
%


\def\TabTopTotalLHC{
\begin{table}[tb]
\centering
\caption{
Integrated hadronic cross section for $t\bar{t}$ production 
at the LHC,
at NLO QED in different production subprocesses,  without and
with cuts. 
\label{tab:top_sigtot_LHC}}
\begin{tabular}{
	c @{\quad}
	p{2mm}
	p{1mm}
	r@{}l @{\quad}
	p{1mm}
	r@{}l
	p{2mm}
	p{1mm}
	r@{}l @{\quad}
	p{1mm}
	r@{}l
}
&&&&&&&&\\[-2mm]
\hline\hline
&&&&&&&&\\[-3mm]
Process
	&& \multicolumn{6}{c}{$\sigma_\mathrm{tot}$ without cuts [pb]}  
	&& \multicolumn{6}{c}{$\sigma_\mathrm{tot}$ with cuts [pb]}  
	\\[2mm]
\cline{3-8} \cline{10-15}
&&&&&&&&\\[-4mm]
	&& \multicolumn{3}{c}{\small Born}
	 & \multicolumn{3}{c}{\small correction}
	&& \multicolumn{3}{c}{\small Born}
	 & \multicolumn{3}{c}{\small correction}
	 \\[1mm]
\hline
&&&&&&&&\\[-4mm]
$u\bar{u}$ && &  34&.25  & &  -1&.41   && &  18&.64   & & -0&.770  \\[1mm]
$d\bar{d}$ && &  21&.61  & &  -0&.228  && &  11&.54   & & -1&.68   \\[1mm]
$s\bar{s}$ && &   4&.682 & &  -0&.0410 && &   2&.253  & & -0&.0304 \\[1mm]
$c\bar{c}$ && &   2&.075 & &  -0&.0762 && &   0&.9630 & & -0&.0446 \\[1mm]
$gg$ && & 407&.8   & &   2&.08   && & 213&.6    & &  0&.524  \\[1mm]
$g\gamma$
   && &    &     & &   4&.45   && &    &      & &  2&.29   \\[1mm]
\hline
&&&&&&&&\\[-4mm]
$pp$ && & 470&.4   & &   4&.78   && & 247&.0    & &  1&.80   \\[1mm]
\hline\hline
&&&&&&&&\\
\end{tabular}
\end{table}
}

\def\TabTopTotalTevatron{
\begin{table}[!ht]
\centering
\caption{
Integrated hadronic cross section for $t\bar{t}$ production 
at the Tevatron,
at NLO QED in different production subprocesses,  without and
with  cuts. 
\label{tab:top_sigtot_Tev}}
\begin{tabular}{
	c @{\quad}
	p{2mm}
	p{2mm}
	r@{}l @{\quad}
	p{1mm}
	r@{}l
	p{2mm}
	p{2mm}
	r@{}l @{\quad}
	p{1mm}
	r@{}l
}
&&&&&&&&\\[-2mm]
\hline\hline
&&&&&&&&\\[-3mm]
Process
	&& \multicolumn{6}{c}{$\sigma_\mathrm{tot}$ without cuts [pb]}  
	&& \multicolumn{6}{c}{$\sigma_\mathrm{tot}$ with cuts [pb]}  
	\\[2mm]
\cline{3-8} \cline{10-15}
&&&&&&&&\\[-4mm]
	&& \multicolumn{3}{c}{\small Born}
	 & \multicolumn{3}{c}{\small correction}
	&& \multicolumn{3}{c}{\small Born}
	 & \multicolumn{3}{c}{\small correction}
	 \\[1mm]
\hline
&&&&&&&&\\[-4mm]
$u\bar{u}$ && & 3&.411                  & & -0&.117
   && & 3&.189                  & & -0&.118 \\[1mm]
$d\bar{d}$ && & 0&.5855                 & & -2&.89$\times$$10^{-3}$
   && & 0&.5432                 & & -2&.91$\times$$10^{-3}$ \\[1mm]
$s\bar{s}$ && & 8&.063$\times$$10^{-3}$ & & -1&.21$\times$$10^{-5}$
   && & 7&.343$\times$$10^{-3}$ & & -1&.79$\times$$10^{-5}$ \\[1mm]
$c\bar{c}$ && & 2&.044$\times$$10^{-3}$ & & -5&.06$\times$$10^{-5}$
   && & 1&.857$\times$$10^{-3}$ & & -5&.00$\times$$10^{-5}$ \\[1mm]
$gg$ && & 0&.4128                 & &  3&.17$\times$$10^{-3}$
   && & 0&.3803                 & &  2&.69$\times$$10^{-3}$ \\[1mm]
$g\gamma$
   && &  &                      & &  0&.0154
   && &  &                      & &  0&.0143 \\[1mm]
\hline
&&&&&&&&\\[-4mm]
$p\bar{p}$ && & 4&.420  & & -0&.102   && & 4&.121 & & -0&.104 \\[1mm]
\hline\hline
&&&&&&&&\\
\end{tabular}
\end{table}
}


\thispagestyle{empty}
\setcounter{page}{0}
\def\thefootnote{\fnsymbol{footnote}}

{\textwidth 15cm

\begin{flushright}
MPP-2007-106\\
hep-ph/0708.1697 \\
\end{flushright}

\vspace{2cm}

\begin{center}

{\Large\sc {\bf NLO QED contributions to top-pair production \\[0.3cm]
                at hadron colliders}}

\vspace{2cm}

\sc{W. Hollik} \rm  ~and \sc{M. Koll\'ar}\rm
\vspace{1cm}

     Max-Planck-Institut f\"ur Physik \\
     (Werner-Heisenberg-Institut)\\
     D-80805 M\"unchen, Germany

\end{center}

\vspace*{2cm}

\begin{abstract}
\noindent
Electroweak one-loop calculations for production of top-quark pairs
at colliders are completed by providing the missing QED type
contributions from real and virtual photons, 
where also effects from interference between QED and QCD contributions
have to be taken into account. Moreover, 
photon-induced
$t \bar{t}$ production is included as another partonic channel.
\end{abstract}

}
\def\thefootnote{\arabic{footnote}}
\setcounter{footnote}{0}

\newpage

\section{Introduction}

\noindent
Experimental investigations of the top quark at the Tevatron have 
significantly contributed to precision tests of the Standard Model (SM) 
since the top  discovery in 1995~\cite{Abe:1995hr,Abachi:1995iq}. 
The top quark mass is an important parameter within the SM and its precise
knowledge is an essential ingredient to constrain the mass of the Higgs 
boson~\cite{Group:2006mx}. Besides the top mass, the measurement of the 
top-pair production cross section is an important test of the SM, and 
possible observation of deviations from the SM predictions could indicate 
new, non-standard,  contributions. Moreover, precise knowledge of the SM 
processes as a main source of background is crucial in direct searches 
for potential new physics beyond the SM.

At the Tevatron, the dominant production mechanism is the annihilation 
of quark-antiquark pairs $ q+\overline{q} \rightarrow t+\overline{t},$
wheras at LHC energies,  $\ttbar$ production proceeds mainly through 
gluon fusion, $g+g\rightarrow t+\overline{t}$. In lowest order, the 
$\ttbar$ production cross section in hadronic collisions is of $\Oass$ 
and was calculated in 
\cite{Gluck:1977zm,*Combridge:1978kx,*Babcock:1977fi,*Hagiwara:1978hw,
*Jones:1977wu,*Georgi:1978kx}. 
The corresponding lowest-order electroweak contributions of $\Oas$ to the 
Drell-Yan annihilation process via $\gamma$-~and $Z$-exchange are very 
small, contributing less than 1\% at the partonic level~\cite{Baur:1989qt}, 
and are thus negligible.
Accordingly, the main higher order contributions arise from QCD. Cross 
sections and distributions including QCD effects of $\Oasc$ were computed 
in \cite{Nason:1987xz,*Nason:1989zy,Beenakker:1988bq,*Beenakker:1990ma}, 
and an inspection of the QCD effects close to the production threshold was 
performed in~\cite{Fadin:1987wz,*Fadin:1990wx}. Including the resummation 
of large logarithmic QCD contributions in the threshold region improves 
the perturbative calculation and was done in 
\cite{Berger:1996ad,*Berger:1997gz,Catani:1996dj,*Catani:1996yz,
Kidonakis:1995wz,Laenen:1991af,*Laenen:1993xr,Bonciani:1998vc}. 
The prediction for the $\ttbar$ production cross section currently used at 
the Tevatron is based on the studies in \cite{Cacciari:2003fi}, which 
include the next-to-leading-order (NLO) contributions and the resummation 
of soft logarithms (NLL). In \cite{Kidonakis:2003qe}, also the 
next-to-next-to-leading-order (NNLO) soft-gluon corrections were taken
into account, extended to NNNLO in~\cite{Kidonakis:2005kz}.

From the electroweak (EW) side, the EW one-loop corrections to the 
QCD-based lowest order calculations, which are of $\Oaass$, were 
investigated first in~\cite{Beenakker:1993yr} for the subclass of the 
infrared-free non-photonic contributions, i.e.~those loop contributions 
without virtual photons. They are of special interest due to the large 
Yukawa coupling of the top quark to the Higgs boson. However, they have 
little impact within the SM, about $1\%$ of the lowest-order cross section 
for the Tevatron, and not more than $3\%$ for the LHC 
\cite{Beenakker:1993yr,Kao:1997bs}. 
In these calculations contributions including the interference of QCD and 
EW interactions were neglected. A study of the non-photonic EW corrections 
with the gluon--$Z$ interference effects was done more recently
in~\cite{Kuhn:2005it,Moretti:2006nf,Bernreuther:2006vg,Kuhn:2006vh}.

Still, a subset of the full EW corrections, corresponding to the QED 
corrections with real and virtual photons, was not included in the previous 
calculations. In this paper we close this gap and present the calculation 
of the missing QED subset, thus making the SM prediction at the one-loop 
level complete.

It is worth to mention also several studies within specific extensions of 
the SM,  comprising calculations of the Yukawa one-loop corrections 
within the General 2-Higgs-Doublet Model (G2HDM) for Tevatron 
\cite{Stange:1993td} and  LHC \cite{Zhou:1996dx}. Also, the SUSY-QCD $\Oas$ 
contributions were calculated for Tevatron 
\cite{Alam:1996mh,*Sullivan:1996ry,*Li:1996jf,*Li:1995fj,Kim:1996nz}
and LHC \cite{Zhou:1997fw}, and for both~\cite{Berge:2007dz}.
The SUSY-EW corrections have been examined: partial calculations relevant 
for the Tevatron were done in \cite{Kim:1996nz}, and a more complete 
description at $\Oaass$ within the G2HDM and MSSM with numerical results 
for Tevatron and LHC, was presented in~\cite{Hollik:1997hm, Kao:1999kj}.

In the following, we provide the QED corrections to top pair production
and also the effects arising from interference of QCD and QED interactions 
that occurs at one-loop order.
Moreover, at this order the distribution of photons inside the proton has 
to be taken into account, adding photon-induced top production as another 
partonic channel at NLO. In the end, we present numerical results for both 
Tevatron and LHC.

Although our calculations are performed in the frame of the SM,
they are also valid for extensions of the SM, as e.g.~the MSSM and G2HDM.

\section{Structure of the QED contributions}

The QED contributions can be treated as a separate subclass at the
electroweak one-loop level.
They consist of virtual and real photon contributions, 
according to the topology of photonic insertions in the lowest-order graphs. 
Both real and virtual photon terms have to be combined in order to
obtain a consistent, infrared (IR) finite result.

\subsection{Virtual corrections}

The virtual QED corrections consist of loop contributions with virtual
photons. They can be described by the matrix elements
$\delta{\cal M}_a$, $a=gg,q\bar{q}$, for both partonic subprocesses
separately. Contracting these quantities with the Born matrix elements
${\cal M}^a_B$ yields the one-loop contribution to the
differential cross sections at the partonic level 
of the order of $\Oaass$, after spin and colour summation,
\begin{equation}\label{parton}
\frac{d\hat \sigma_a^{\rm (1-loop)} } 
{d\hat t} (\hat t,\hat s)
= \frac{1}{16 \pi \hat s^2}\, \cdot
2\, {\rm Re}  \; \overline{\sum}(\delta{\cal M}_a \cdot
{\cal M}_B^{a*})\; ,
\end{equation}
where $\hat t$ and $\hat s$ 
are the usual Mandelstam variables.
The explicit expressions of ${\cal M}^a_B$ are given in 
\cite{Beenakker:1993yr}, where the non-photonic electroweak corrections have 
been studied.
Throughout this paper we closely follow the notation of 
\cite{Beenakker:1993yr}.

\begin{figure}[!tb]
 \centering
  {\includegraphics[angle=0, width=0.8\textwidth]{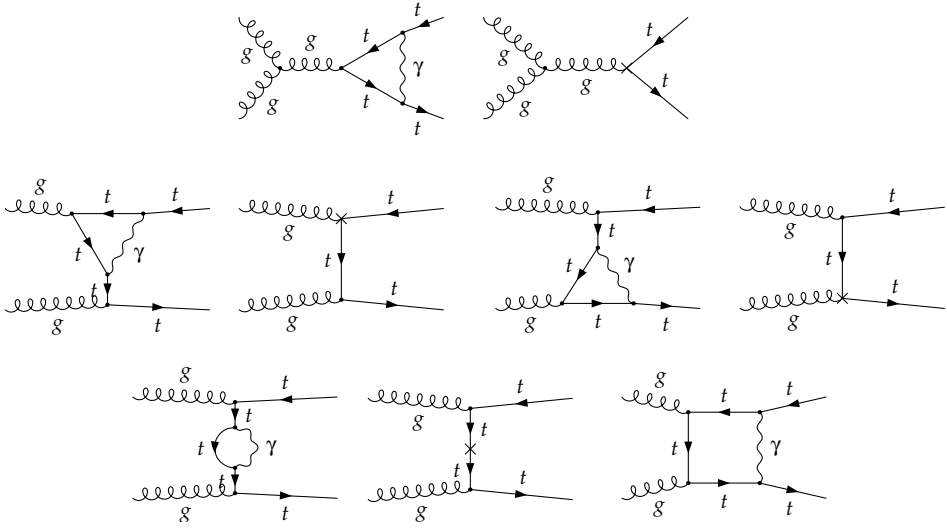}}
\caption{Virtual QED $\Oaass$ contributions to gluon fusion 
($u$-channel diagrams are not explicitly shown). Crossed 
lines and vertices denote counter term insertions.
\label{f:gfusion}}
\end{figure}

\begin{figure}[!tb]
 \centering
  {\includegraphics[angle=0, width=0.8\textwidth]{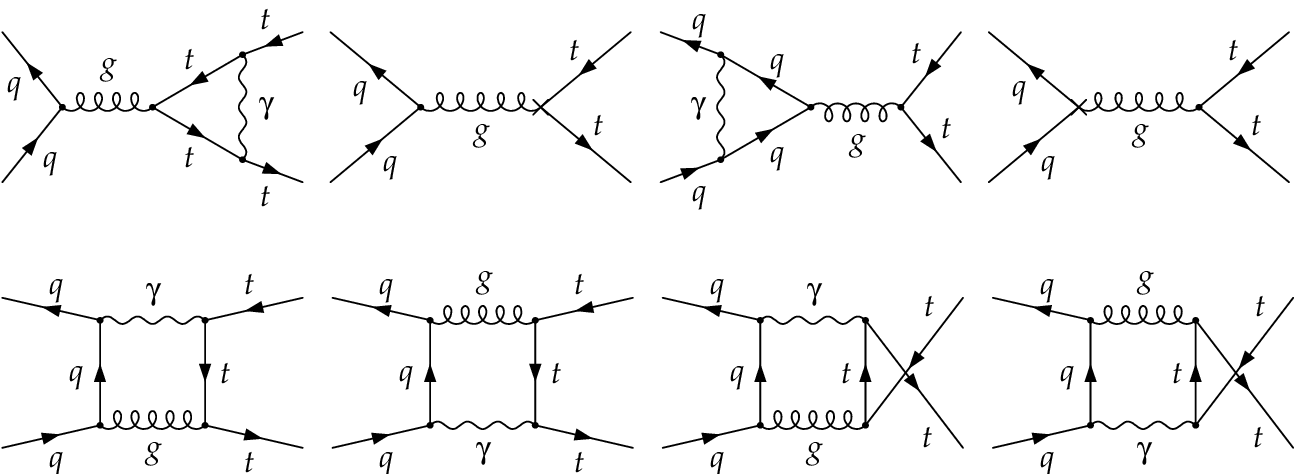}}
\caption{Virtual QED $\Oaass$ contributions to $\qqbar$ annihilation.
        \label{f:qq}}
\end{figure}

The virtual QED corrections of $\Oaass$, contributing to $\delta{\cal M}_a$,
can be classified according to self-energy,
vertex, and box corrections, 
depicted by Feynman diagrams 
in~Figs.~\ref{f:gfusion} and \ref{f:qq} for the two partonic 
processes of $q\bar{q}$ annihilation and gluon fusion.
They were treated with the help of the 
{\em FeynArts}~\cite{Kublbeck:1990xc, *Hahn:2000kx, *Hahn:2001rv},
{\em FormCalc}~\cite{Hahn:1998yk}, and
{\em LoopTools}~\cite{Hahn:1999mt, *Hahn:2000jm, *Hahn:2006qw} packages,
based on techniques from \cite{'tHooft:1978xw, Passarino:1978jh},
which were further refined for 4-point  integrals 
in~\cite{Beenakker:1988jr,Denner:1991qq}.
The analytical expressions for the matrix elements
are close to those 
in~\cite{Beenakker:1993yr}
and \cite{Beenakker:1991ca}.

The whole set of QED loop diagrams
is gauge invariant and UV-finite after taking into account
the counter terms for the $g \qqbar$- vertex, $g \ttbar$-vertex 
and top quark self-energy.
UV singularities in the sum of vertex functions 
and corresponding counter terms 
with quark field and mass renormalization constants cancel,
hence, no coupling constant renormalization is needed.

To obtain finite vertices and propagators, it is thus sufficient to 
perform field and mass renormalization for the quarks.
In case of top quarks, the substitution 
\begin{eqnarray}  
\Psi_{t} & \rightarrow & \left( 1+ \frac{1}{2} \delta Z_{t} 
\right) \Psi_{t} \; ,\nonumber \\  
m_t & \rightarrow & \; \; m_t - \delta m_t \; ,  
\end{eqnarray}
yields the counter term for the
$g \ttbar$-vertex, $\delta\Lambda_\mu$, and for the top self-energy,
$\delta\Sigma$, as follows: \vspace{3mm} \newline
\begin{minipage}[l]{0.3\linewidth}
\hspace{.18\linewidth}\includegraphics[height=3cm]{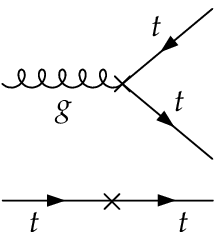}
\end{minipage}
\begin{minipage}[r]{0.69\linewidth}
\begin{eqnarray} 
i\delta\Lambda_\mu & = & -i g_s T^c\gamma_\mu \,\delta Z_t \; , \\[1cm] 
\label{eq:top_counter} 
i\delta\Sigma & = & i \left( p\!\!\!/ \,\delta Z_t - m_t \,\delta Z_t + \delta  
m_t \right) \; .
\end{eqnarray}
\end{minipage}

\vspace{4mm}\noindent 
The field renormalization constant $\delta Z_t$ as well as
the mass counter term $\delta m_t$ are
fixed by renormalization conditions, for which we 
choose the on-shell scheme.
They are imposed on the renormalized top quark self-energy 
$\hat \Sigma$ = $\Sigma$ + $\delta\Sigma$,
with 
\begin{equation} 
\Sigma (p) = \frac{\alpha}{4\pi} \left[ 
p\!\!\!/ \, \Sigma_V (p^2) + m_t \Sigma_S (p^2) \right] \; , 
\end{equation}
corresponding to the QED-like part of the unrenormalized
top quark self-energy,
\vspace{5mm} \newline
\begin{minipage}[l]{0.3\linewidth}
\hspace{.24\linewidth}\includegraphics[height=2.cm]{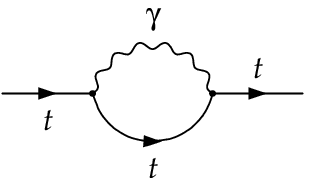}
\end{minipage}
\begin{minipage}[r]{0.69\linewidth}
\begin{eqnarray} 
, \\[1mm] 
\nonumber 
\end{eqnarray}
\end{minipage}
\vspace{2mm}
\newline
in the following way:
\begin{enumerate}
\item[(i)]{The pole of the top
quark propagator
is kept at $m_t$ and thus defines the on-shell mass:
\begin{equation} 
{\rm Re}\, \hat \Sigma \left( p\!\!\!/ \, = m_t \right) = 0 \; , 
\quad {\rm yielding} \quad
\frac{\delta m_t}{m_t}  = -(\Sigma_V+\Sigma_S)\big|_{p^2=m_t^2} \; . 
\end{equation}
}
\item[(ii)]
{The residue of the top quark propagator is unity, 
yielding the field renormalization constant by (real parts only)
\begin{equation} 
\label{eq:top_dZ} 
\delta Z_t = -\Sigma_V(m_t^2)-2m_t^2\, \frac{\partial}{\partial p^2}  
(\Sigma_V+\Sigma_S)\big|_{p^2=m_t^2} \; . 
\end{equation}
}
\end{enumerate}

The renormalization constants for initial-state quarks $(q)$ are determined
analogously, substituting
$m_t \rightarrow m_q$. 
To obtain the counter terms for the initial state
$g \qqbar$ vertices, only the renormalization of the quark fields is 
necessary. Light quark masses are only kept where they are necessary
to regularize collinear divergences, which appear as double logarithms 
$\ln^2 (\hat{s}/m_q^2)$
and single logarithms $\ln (\hat{s}/m_q^2)$.

As a consequence of the null photon mass, the virtual QED corrections are
IR divergent. The photonic IR singularities can be regularized by
introducing a fictitious photon mass $\lambda$.

A specific peculiarity of the QED corrections are the 
$\Oaas$ box contributions shown in Fig.~\ref{f:qq}, which 
contain besides photons also gluons in the loop. As a consequence, 
further IR singularities 
related to the gluons emerge from the loop integrals. 
Since the gluons appear quite similar to the photons in the box graphs,
it is possible to
perform the regularization by a gluon mass as well. 
For simplicity, we use the 
same regularization parameter $\lambda$.

\subsection{Real corrections}

\begin{figure}[!tb]
 \centering
  {\includegraphics[angle=0, width=0.8\textwidth]{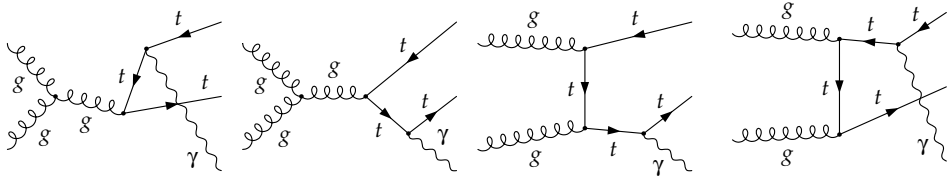}}
\caption{Real QED $\Oaass$ contributions of photon bremsstrahlung
to the $gg$ fusion (u-channel diagrams are not explicitly shown).
        \label{f:gg23}}
\end{figure}

\begin{figure}[!tb]
 \centering
  {\includegraphics[angle=0, width=0.8\textwidth]{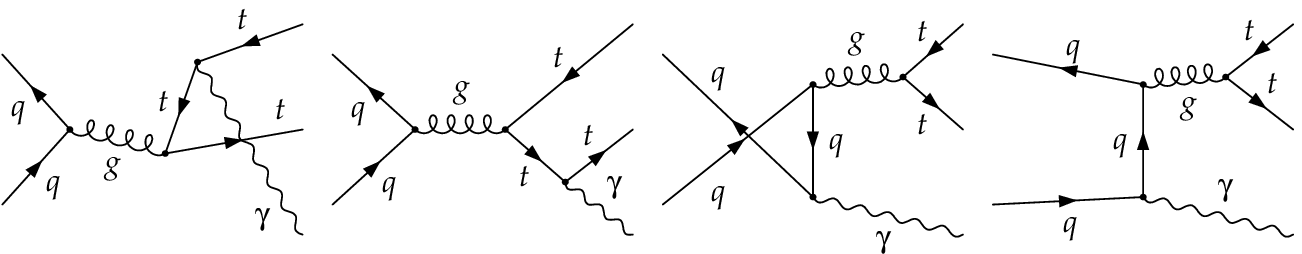}}
\caption{Real QED $\Oaass$ contributions of photon bremsstrahlung to the 
$\qqbar$ annihilation.
        \label{f:qq23}}
\end{figure}

According to the Bloch-Nordsieck theorem \cite{Bloch:1937pw}, the IR singularities
in the virtual corrections cancel against their counterparts from the real
photon contributions after integration over the photon phase space. 
Therefore we have to include all contributions of the real 
photon radiation off the external particles to obtain 
an IR finite result. The corresponding diagrams are shown in 
Figs.~\ref{f:gg23} and \ref{f:qq23}.

\begin{figure}[tb]
 \centering
  {\includegraphics[angle=0, width=0.8\textwidth]{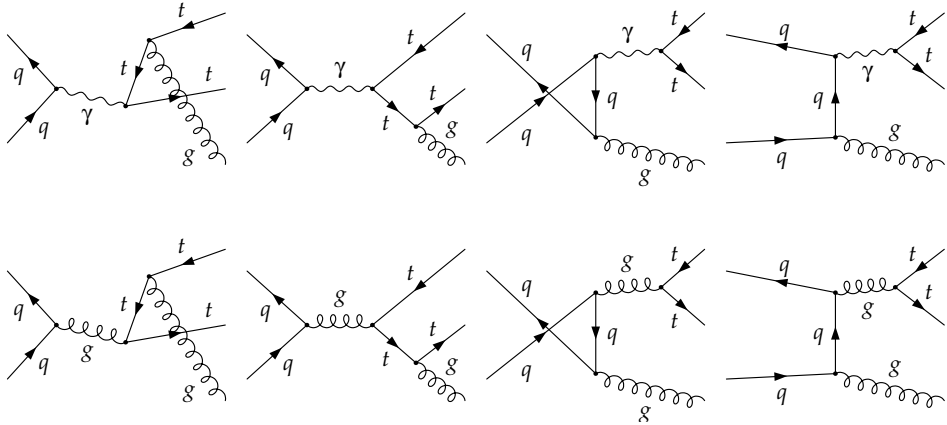}}
\caption{Gluon bremsstrahlung from  QED-mediated
(upper row) and  QCD-mediated (second row) Born diagrams,
contributing at $\Oaass$ through interference.
        \label{f:gluonbr}}
\end{figure}

Moreover, we have to include also gluon bremsstrahlung to compensate the IR
singularities related to the gluons in the box graphs of Fig.~\ref{f:qq}.
They consist of two types of diagrams:
gluon radiation off the QED-mediated and 
off the QCD-mediated $\qqbar$ annihilation, 
as depicted in Fig.~\ref{f:gluonbr}.

\begin{figure}[tb]
 \centering
  {\includegraphics[angle=0, width=0.45\textwidth]{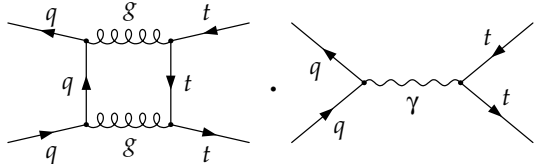}}
\caption{$\Oaass$ contribution to  $\qqbar$ annihilation via 
QCD box diagrams (crossed diagram not explicitly shown) interfering
with the QED Born diagram.
        \label{f:qcdbox}}
\end{figure}

At  $\Oaass$, it is the interference of these two 
classes of diagrams that is required, yielding a new type of QED--QCD 
interference. 
Still, not all of the interference terms contribute.
Owing to the color structure,
only the interference of the initial and final
state gluon radiation graphs is non-zero,
yielding the structure required to cancel the IR singular parts  
in the box corrections of Fig.~\ref{f:qq}.
Nevertheless, the cancellation is not yet complete.
The missing piece is the pure QCD box correction 
interfering with the QED $\qqbar$ annihilation Born level 
diagram, as displayed in Fig.~\ref{f:qcdbox}, 
which gives another non-zero contribution
of the same order. 
Only after combining all these various parts the
$\Oaass$ result is IR finite.

\begin{figure}[tb]
 \centering
  {\includegraphics[angle=0, width=0.4\textwidth]{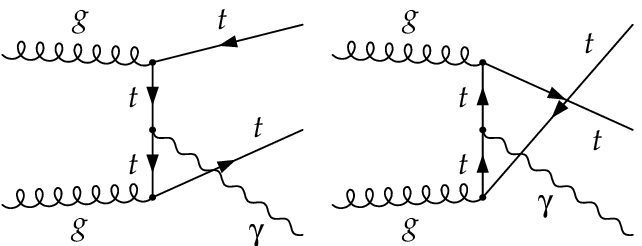}}
\caption{IR-finite $\Oaass$ bremsstrahlung contributions. 
        \label{f:inner}}
\end{figure}

In Fig.~\ref{f:qcdbox}, only the photon-mediated Born-level diagram is shown.
In principle, also the interference of the QCD box
and $Z$-boson exchange tree-level diagram has to be taken into account. 
This contribution belongs to the IR-singular gluon--$Z$ corrections, which
also contain the gluon--$Z$ box graphs and
gluon bremsstrahlung off
$Z$-mediated tree-level diagrams. 
The IR-singular structure of these contributions is simplified by the fact
that there are no IR-singularities related to the $Z$-boson. 
The gluon--$Z$ interference effects were neglected in the original
study of non-photonic EW corrections performed in \cite{Beenakker:1993yr}. 
They have been investigated recently in 
\cite{Kuhn:2005it,Moretti:2006nf,Bernreuther:2006vg,Kuhn:2006vh}.

For completeness of the NLO QED effects,  
photon radiation off the off-shell top quarks in the $gg$
fusion subprocess (Fig.~\ref{f:inner}) has to be considered as well.
These effects are, however, one order of magnitude
smaller than the other terms and hence 
are less important for numerical studies.

Technically, for the phase space integration of real photon/gluon radiation,
we apply the phase space slicing method (see appendix) 
taking advantage of its universality in handling both inclusive and
non-inclusive quantities. The dipole subtraction method~\cite{Dittmaier:1999mb},
originally proposed for QCD~\cite{Catani:1996jh,*Catani:1996vz}, 
was used to verify numerical results obtained with the slicing method at the
partonic level.

\subsection{Photon-induced {\boldmath{$\ttbar$ production}}}
\label{sec:phot_top}

\begin{figure}[tb]
 \centering
 \includegraphics[angle=0, width=0.42\textwidth]{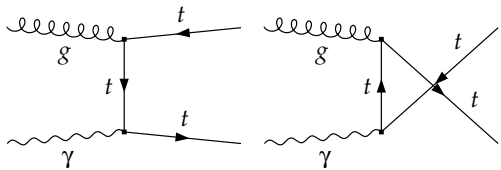}
\caption{Feynman diagrams for photon induced $\ttbar$ production
at lowest order.
   \label{f:gga}}
   \end{figure}

In addition to the previously mentioned NLO QED contributions we also
have to inspect the photon-induced production channels. These comprise
at lowest order the gluon--photon fusion amplitudes illustrated
in Fig.~\ref{f:gga}.

In general, photon-induced partonic processes vanish at the hadronic
level unless the NLO QED effects are taken into account. A direct consequence 
of including these effects into the evolution of parton distribution functions 
(PDFs) is the non-zero photon density in the proton, which leads to 
photon-induced contributions at the hadronic level by
convoluting the photon-induced partonic cross sections with the PDFs at NLO QED.
Since the photon distribution function 
is of order $\alpha$
they are formally not of the same overall order as the other NLO 
QED contributions. Numerically, however, they turn out to be
sizeable, and we therefore include them in our discussion.

As the PDFs at NLO QED have become available only recently \cite{Martin:2004dh},
the photon-induced hadronic processes have not yet been investigated.
Here we present the first study of these effects on the top pair production.

\section{Hadronic cross section for 
 {\boldmath{$pp,\ppbar \rightarrow \ttbar X$}}}

For obtaining the hadronic cross section we have to convolute the
various partonic cross sections with the corresponding parton densities
and sum over all contributing channels,
adding up contributions of the non-radiative and radiative processes. 
As already mentioned, only the sum of all virtual and
real corrections is IR finite. Final step is the
factorization of the remaining mass singularities.

\subsection{Mass factorization}

The mass-singular logarithmic terms proportional to $\ln m_q$ are not
canceled in the sum of virtual and real corrections. They originate from
collinear photon emission off the incoming light quarks. In analogy
to the factorization of collinear gluon contributions, they have to be
absorbed into the parton densities.

This can be formally achieved by replacing the bare quark distributions
$q_i(x)$ for each flavor 
by the appropriate scale dependent distributions $q_i(x,Q^2)$
in the following way, according to \cite{Baur:1998kt} 
(with $m_i = m_{q_i}$),
\begin{eqnarray} 
q_i(x,Q^2) &=& q_i(x) \nonumber \\  
&+& \frac{\alpha}{\pi}Q_i^2 q_i(x) \Bigg\{ 1 - \ln  
\delta_s - \ln^2 \delta_s  + \left( \ln \delta_s + \frac{3}{4} \right)  
\ln \left( \frac{Q^2}{m^2_i} \right) \nonumber \\ 
&&-\frac{1}{4}\lambda_{FC}f_{v+s} \Bigg\} \nonumber \\ 
&+& \int_x^{1-\delta_s} \frac{dz}{z}q_i\left(\frac{x}{z}\right) 
\frac{\alpha}{2\pi} Q^2_i \Bigg\{ \frac{1+z^2}{1-z} \ln \left(  
\frac{Q^2}{m_i^2} \frac{1}{(1-z)^2} \right) \nonumber \\ 
&&-  \frac{1+z^2}{1-z} + 
\lambda_{FC}f_c \Bigg\} \; , 
\end{eqnarray}
involving a soft-energy cut $\delta_s$ and the functions
\begin{equation} 
f_{v+s} = 9+\frac{2\pi^2}{3}+3\ln \delta_s - 2\ln^2 \delta_s \; , 
\end{equation}

\begin{equation} 
f_c = \frac{1+z^2}{1-z}\ln \left(\frac{1-z}{z} \right) -\frac{3}{2(1-z)} 
+2z +3 \; . 
\end{equation}
The expressions for the PDFs are given in both DIS and $\overline{\rm MS}$
factorization schemes, which corresponds to $\lambda_{FC}=1$ and
$\lambda_{FC}=0$, respectively.

For a consistent treatment of the collinear singularities at $\Oa$, it is
necessary to use an appropriate set of PDFs that was extracted from the data
and evolved by DGLAP equations with the NLO QED effects included. Otherwise
it would lead to an overestimation of the scale dependence. We
use the PDFs from the MRST collaboration \cite{Martin:2004dh} which were
determined at NLO QCD and NLO QED.
The authors do not explicitly state which factorization scheme is relevant
for NLO QED.
We follow the reasoning given in \cite{Diener:2005me} and use the DIS scheme
in our calculation. For the numerical evaluation 
the factorization scale is set to $Q = \mu_F = 2m_t$.
After performing the factorization of mass singularities, the results become
free of the quark-mass logarithms.
The scale dependence cannot be checked in a consistent
way owing to the NLO QCD effects in the parton densities, as these are not
included in our calculation. For this reason we do not present a study of
the scale dependence.

\subsection{Integrated hadronic cross sections}

The observable hadronic cross section is obtained by convoluting the short
distance partonic cross sections $\hat\sigma_{gg}$, $\hat\sigma_{\qqbar}$,
$\hat\sigma_{\gamma g}$
with the universal parton distribution functions
for quarks, gluons, and photons. 
For colliding hadrons $A$ and $B$ carrying 
the momenta $P_A$ and $P_B$, with
$S = (P_A+P_B)^2$, the hadronic
cross section can be expressed as
\begin{equation} 
\label{eq:hadrtotsig} 
\sigma(S) = \int_{\frac{4 m_t^2}{S}}^1 d\tau  
\left[ \sum_i \, {\cal L}^{AB}_{q_i \bar{q_i}}(\tau) \; 
\hat  \sigma_{q_i\bar{q_i}} (\hat s)  
+{\cal L}^{AB}_{gg}(\tau) \; \hat \sigma_{gg}(\hat s)
+{\cal L}^{AB}_{\gamma g}(\tau) \; \hat \sigma_{\gamma g}(\hat s)
\right] \; , 
\end{equation}
with $\tau= \hat s/S$,
and $\hat s$ the partonic center-of-mass
energy squared. 
The partonic cross sections $\hat \sigma_{\qqbar}$
and  $\hat \sigma_{gg}(\hat s)$
include the virtual photon and gluon loop contributions 
as well as the  real photon and gluon bremsstrahlung
terms, as described in section 2; 
the photon-induced partonic cross section $\hat \sigma_{\gamma g}$
is of lowest order.

The parton luminosities are defined as follows,
\begin{equation} 
\label{eq:hadrlumin} 
{\cal L}^{AB}_{mn}(\tau) = \frac{1}{1+\delta_{mn}}   
\int_{\tau}^1 \frac{d x}{x} 
\left[ \Phi_{m/A}(x,\mu_F)\Phi_{n/B}(\frac{\tau}{x},\mu_F)   
+(1 \leftrightarrow 2)\right] \; ,  
\end{equation}
with the parton distributions inside $A$, $\Phi_{m/A} = q_i, g, \gamma $,
and the factorization scale, $\mu_F$.

\subsection{Differential hadronic cross sections}

In addition to the integrated hadronic cross section it is convenient
to define hadronic cross sections differential in one or more parameters.
Typically, the variables are chosen to be Lorentz invariant quantities or
quantities with simple transformation properties. In our study we consider
the differential cross sections with respect to the invariant mass of
the $\ttbar$ pair (inclusive) and with respect to the transverse momentum of
top quark.

The invariant mass distribution of the hadronic cross
section has the following form,
\begin{equation}
\label{eq:hadrinvmass} 
\frac{d\,\sigma}{d\sqrt{\hat s}} =  
\frac{2\sqrt{\hat s}}{S}\,
 \sum_{\{m,n\}}\,  {\cal L}^{AB}_{mn} \left( \frac{\hat s}{S} \right)  
      \;  {\hat \sigma}_{mn}(\hat s) \; , 
\end{equation}
where the sum extends over the various partons $m$,$n$  
in the initial state.

The differential hadronic cross section with respect to the transverse 
momentum of the top quark, $\pt$, 
can be written for the partonic $2\to2$ processes as follows,
\begin{equation}\label{eq:hadrpt} 
\frac{d\sigma}{d\pt} =  
\int^1_{\tilde{\tau}_0} 
d\tau\, {\cal L} (\tau)  \,
\frac{\partial {\hat t}}{\partial \pt}
\frac{d\hat\sigma}{d \hat t}(\hat t,\hat s) \; 
\end{equation}
(dropping parton indices and summation).
The lower limit for the
$\tau$-integration is thereby dependent on $\pt$,
\begin{equation} 
\tilde{\tau}_0 = \frac{4(m_t^2+\pt^2)}{S} \; .
\end{equation}

Real photonic/gluonic corrections belong to 
three-particle final states. 
Therefore,
we also need the differential hadronic cross section with
respect to the transverse momentum of the top quark, 
in each of the  $2\to 3$ parton processes. Expressed in terms 
of variables of the parton CMS, it can be written in the 
following way,
\begin{equation} 
\frac{d\sigma}{d\pt} =  \int^1_{\tilde{\tau}_0} d\tau\, {\cal L}(\tau)  
\, \int d k^0_1 \int d k^0_3  \int d \phi_3 \; 
\frac{\partial\! \cos\theta}{\partial \pt} \,
\frac{d\hat\sigma}{d k^0_1\, d k^0_3\, d\phi_3\, d\! \cos\theta}  \; , 
\end{equation}
where $k_1$ and $k_3$ are the 4-momenta of the top quark and the photon(gluon).
$\theta$ is the angle between 
$\vec{k}_1$ and the beam axis, given by $\vec{P}_A$, and
$\phi_3$ is the azimuthal angle of $\vec{k}_3$ with respect to
$\vec{k}_1$ as polar axis.
The threshold for the $\tau$ integration corresponds to
\begin{equation} 
\tilde{\tau}_0 S = \left(\sqrt{m^2_t+\pt^2}+  
\sqrt{(m_t+\lambda)^2+\pt^2}  \right)^2 \, ,
\end{equation}
with the IR mass regulator $\lambda$.

\section{Numerical results}

In the following we present numerical results for the total hadronic cross
section as well as for the distributions with respect
to the invariant mass of the $\ttbar$ pair and the transverse momentum of
the top quark.

For the identification of $\ttbar$ pairs and event reconstruction
it is necessary to apply kinematical cuts, such as cuts to the transverse
momentum and the pseudorapidity $\eta$
of the $t$ and $\bar{t}$.
In the case of the LHC, the cuts applied are as follows,
\begin{equation}  
\pt > 100 \;{\rm GeV}  \quad {\rm and} \quad |\eta| < 2.5 \, . 
\end{equation}
For the Tevatron, the cuts are chosen according to
\begin{equation}  
\pt > \;25 \;{\rm GeV}  \;\quad {\rm and} \quad |\eta| < 2.5 \,.  
\end{equation}

\TabTopTotalLHC

In Table~6.2 and 6.3, we present the numerical results for the integrated
hadronic cross sections at the LHC and at the Tevatron, respectively.
The values of $\sigma$ are listed for the Born level and for the NLO QED 
corrections. The contributions of all production channels are shown separately, 
as well as combined to the total correction.

\TabTopTotalTevatron

At the LHC, the largest correction comes from the photon--gluon production
channel at NLO. It has the same sign as the contribution
to the $gg$ fusion which is the dominant $\ttbar$ production channel at
the LHC. However, the contributions to $\qqbar$ annihilation
have opposite signs which leads to a reduction of the overall NLO QED
correction. In total, the relative correction is about 1\% and is
slightly reduced if the cuts are applied.

At the Tevatron, the largest contributions to the total hadronic cross
section come from the $u\bar u$ subprocess. The photon--gluon subprocess
yields the second largest contribution, but with opposite sign.
In total, the relative correction can amount
to 2.5\%, including cuts.

For illustration of the numerical impact of the NLO QED corrections
on the distributions,
we introduce the relative correction
$\delta$, defined as
\begin{equation} 
\label{eq:delta}  
\delta = \frac{d\sigma_{\textrm{NLO}} - d\sigma_{\rm B}}{d\sigma_{\rm B}} \; , 
\end{equation}
with cross section at NLO, $d\sigma_{\textrm{NLO}}$,
and the Born cross section $d\sigma_{\rm B}$.

\begin{figure}[tb]
 \centering
 \vspace*{2mm}
   \includegraphics[angle=0, width=1\textwidth]{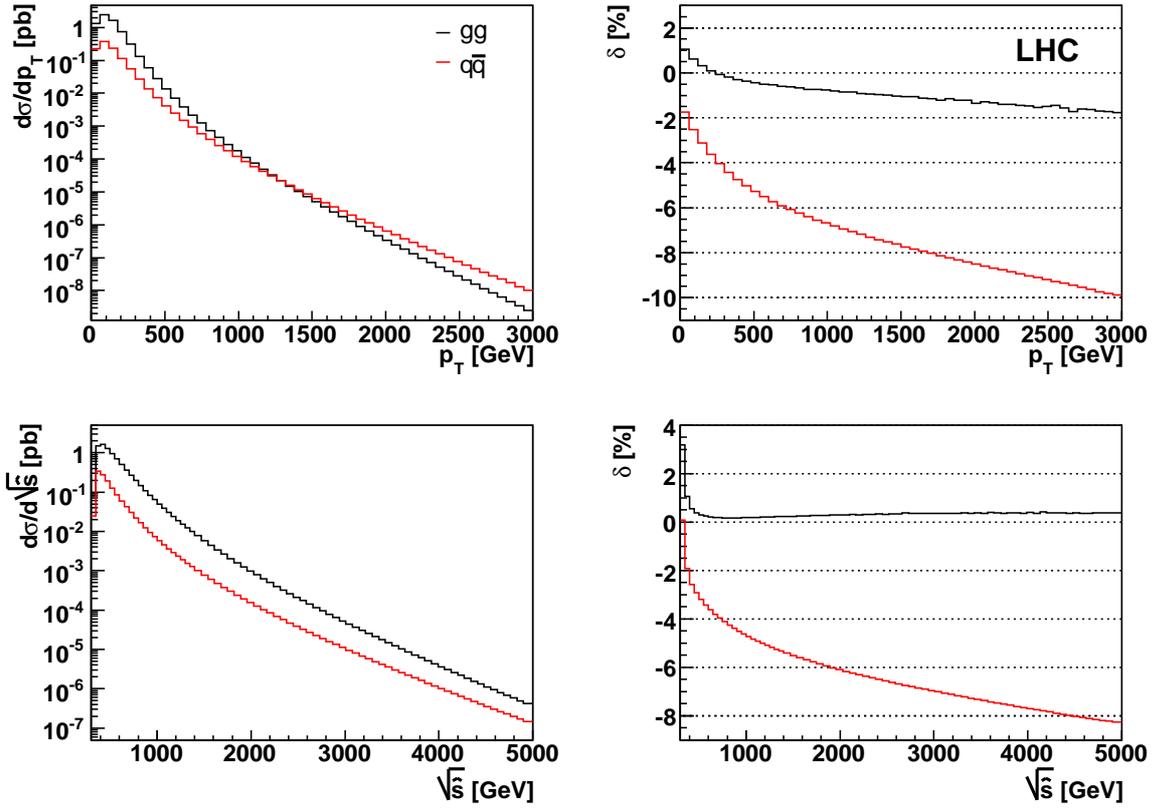}
   \caption{Differential cross sections (left) and
   relative correction $\delta$ (right), as
   functions of the transverse momentum of the top quark (up) and of the
   parton energy (down), at the LHC,
   with no additional cuts.
   \label{fig:L_nocuts}}
\end{figure}

In Fig.~\ref{fig:L_nocuts} the $\pt$ and $\sq$ distributions 
are shown (left), as well as the
relative QED corrections (right), for the $gg$ and $\qqbar$ parton 
channel at the LHC. 
The effect of the NLO QED corrections in the dominant
$gg$ fusion channel is rather small, 
less than 1\% over most of the $\pt$ range and also
over most of the $\sq$ range.
Differently from  the $gg$ channel, 
the NLO contributions for $\qqbar$ annihilation
are negative over the whole $\pt$ and $\sq$ range,
reaching the 5\% level for
$\pt \gtrsim 400$~GeV and $\sq \gtrsim 1200$~GeV. They further grow
in size with increasing $\pt$ and $\sq$ and for very high $\pt$ the 
$\qqbar$ channel starts to dominate over the $gg$ fusion.

\begin{figure}[tb]
 \centering
   \includegraphics[angle=0, width=1\textwidth]{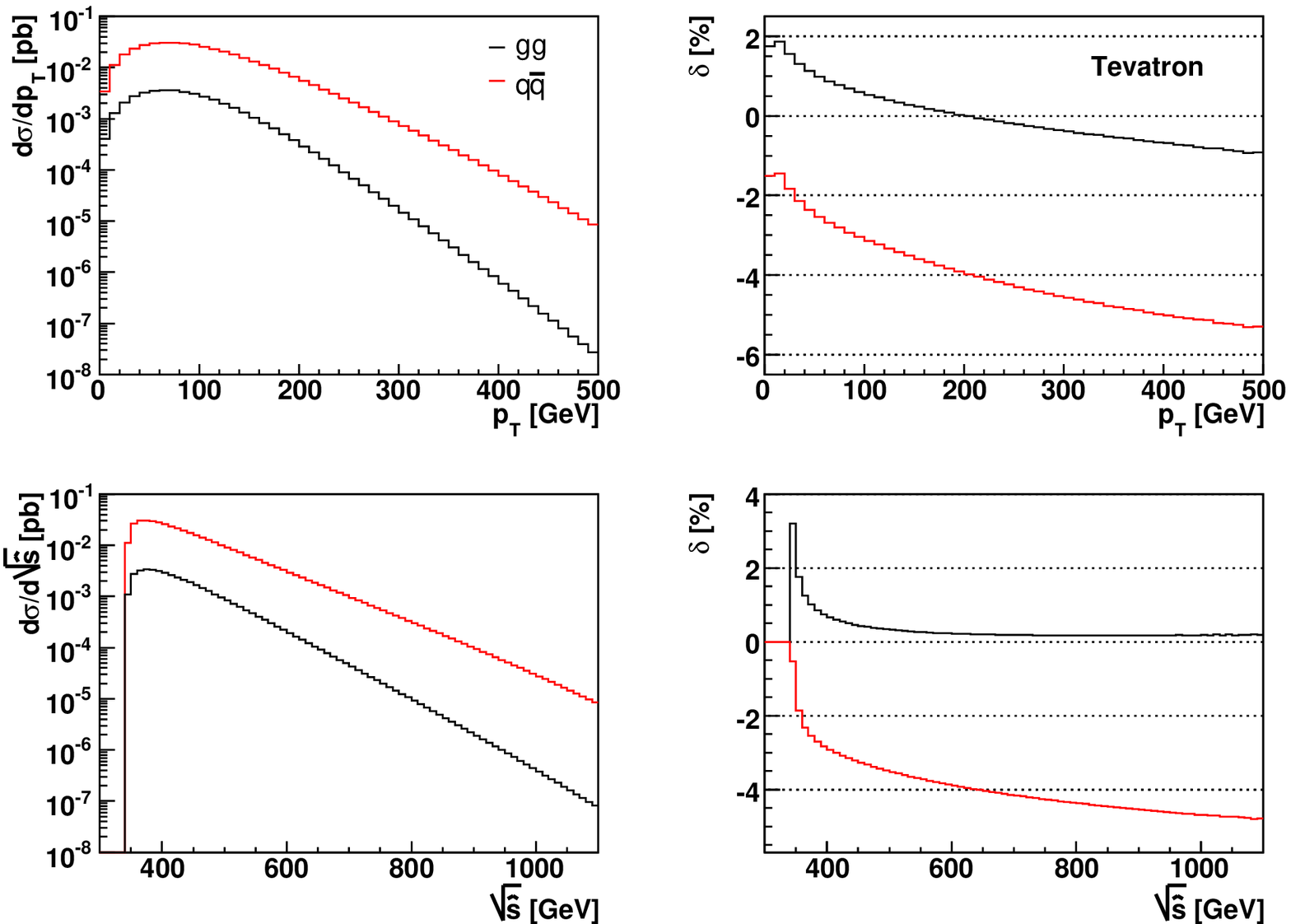}
   \caption{Differential cross sections (left) and
   relative correction $\delta$ (right), as
   functions of the transverse momentum of the top quark (up) and of the
   parton energy (down), at the Tevatron,
   with no additional cuts.
   \label{fig:T_nocuts}}
\end{figure}

In case of the Tevatron [Fig.~\ref{fig:T_nocuts}], the $\qqbar$
annihilation dominates over the $gg$ fusion (left). The impact of $\Oaass$
corrections on both channels is similar to the LHC. Again, in the $gg$
fusion, the relative correction $\delta$ is smaller than 1\% for most of the
$\pt$ and $\sq$ ranges, except for the low $\pt$ and threshold regions
where it reaches about 2\% (up right). In the $\qqbar$ annihilation channel,
the relative corrections are negative and at a few per cent level already near
the threshold. They grow further in size with increasing $\pt$ and $\sq$.
The 5\% level is reached for $\pt \gtrsim 350$~GeV and $\sq \gtrsim 900$~GeV.

\begin{figure}[tb]
\centering
   \includegraphics[angle=0, width=1\textwidth]{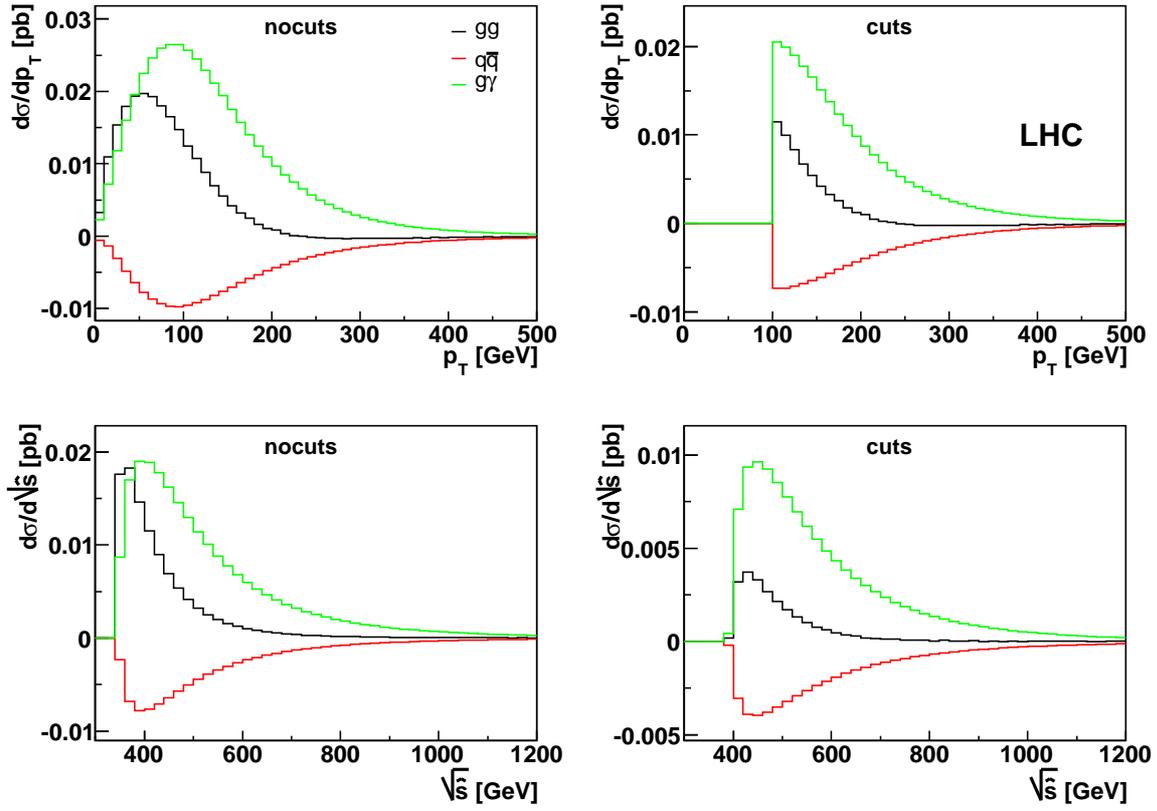}
   \caption{
   NLO QED contributions from the $gg$, $\qqbar$ and $g\gamma$ channels
   at the LHC for the 
   $p_{\rm T}$ and $\sqrt{\hat s}$ distributions,
   including also cuts.
   \label{fig:L_corr}}
\end{figure}

\begin{figure}[tb]
\centering
   \includegraphics[width=1\textwidth]{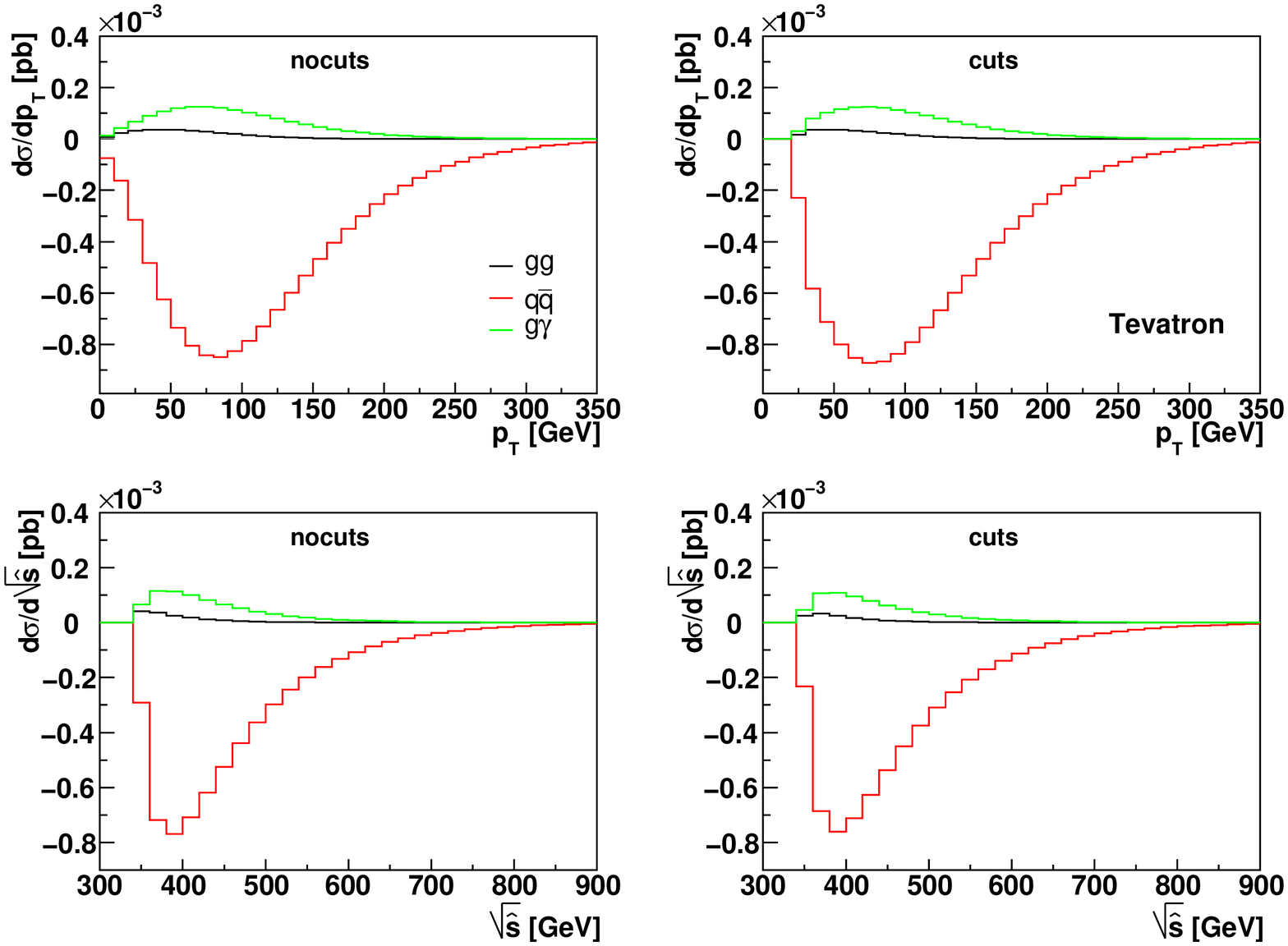}
   \caption{
   NLO QED contributions from the $gg$, $\qqbar$ and $g\gamma$ channels
   at the Tevatron for the $p_{\rm T}$ and $\sqrt{\hat s}$ distributions, 
   including also cuts.
   \label{fig:T_corr}}
\end{figure}

As previously discussed, also the photon-induced processes
represent contributions of NLO in QED, owing to higher order effects
included in the PDFs. In Figs.~\ref{fig:L_corr} and \ref{fig:T_corr}, we
show the photon--gluon contribution to $\ttbar$ production, in 
comparison with the NLO terms in the $gg$ and $\qqbar$
channels. 
For the LHC [Fig.~\ref{fig:L_corr}],
the photon-induced contribution is larger than
the corrections to both Born processes, a consequence of the fact that
the combination of gluon and photon parton densities can become
substantially  large. 
Since $g\gamma$ hadronic cross section is of the same
sign as the NLO contributions to the $gg$ fusion channel,
its presence enhances the size of the overall NLO QED contributions.

The situation is different at the Tevatron [Fig.~\ref{fig:T_corr}],
where the $\qqbar$ annihilation channel dominates. 
Still, the photon-induced contribution is
larger in size than the NLO QED corrections to the $gg$ fusion channel.
However, as a consequence of opposite signs, these two tend to reduce
the contribution from the $\qqbar$ channel. 

\begin{figure}[tb]
 \centering
   \includegraphics[angle=0, width=1\textwidth]{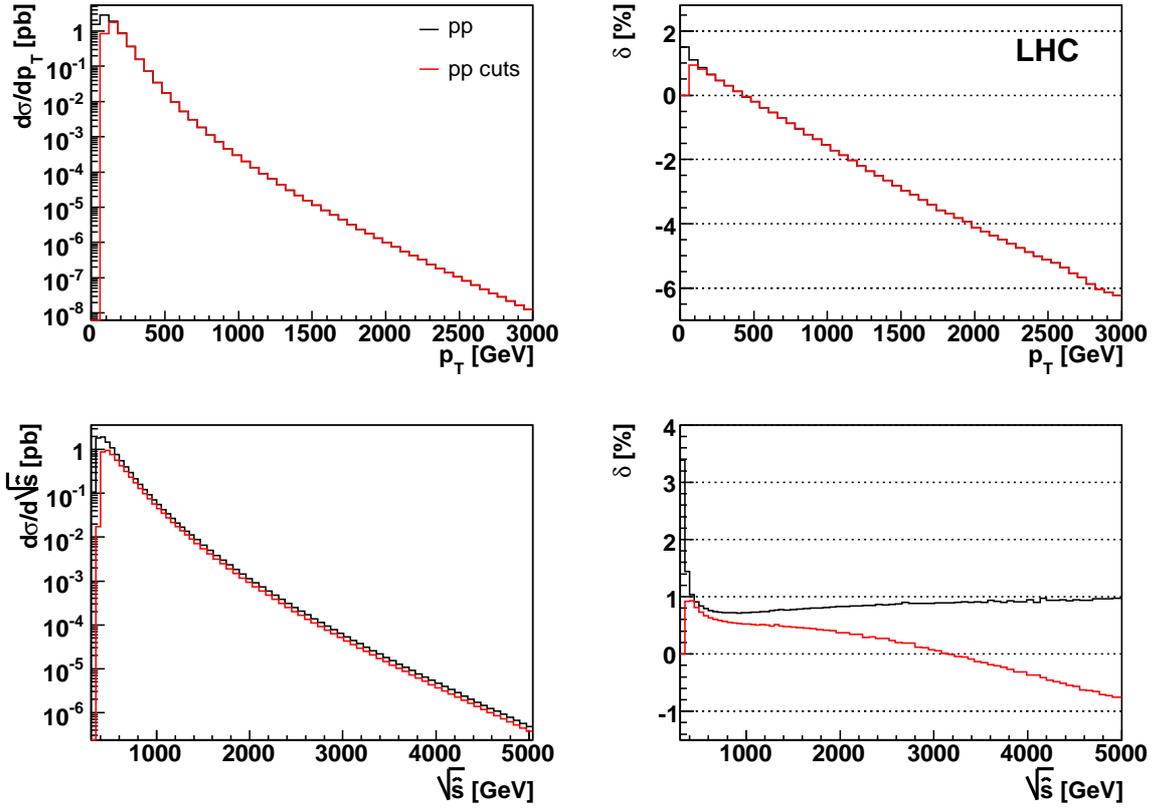}
   \caption{
   Overall NLO QED effects in  $pp$ collisions at the LHC, for the 
   $p_{\rm T}$ distribution (up) and the $\sqrt{\hat s}$ distribution (down),
   without and with application of cuts.
   \label{fig:L_pp}}
\end{figure}

\begin{figure}[tb]
 \centering
   \includegraphics[angle=0, width=1\textwidth]{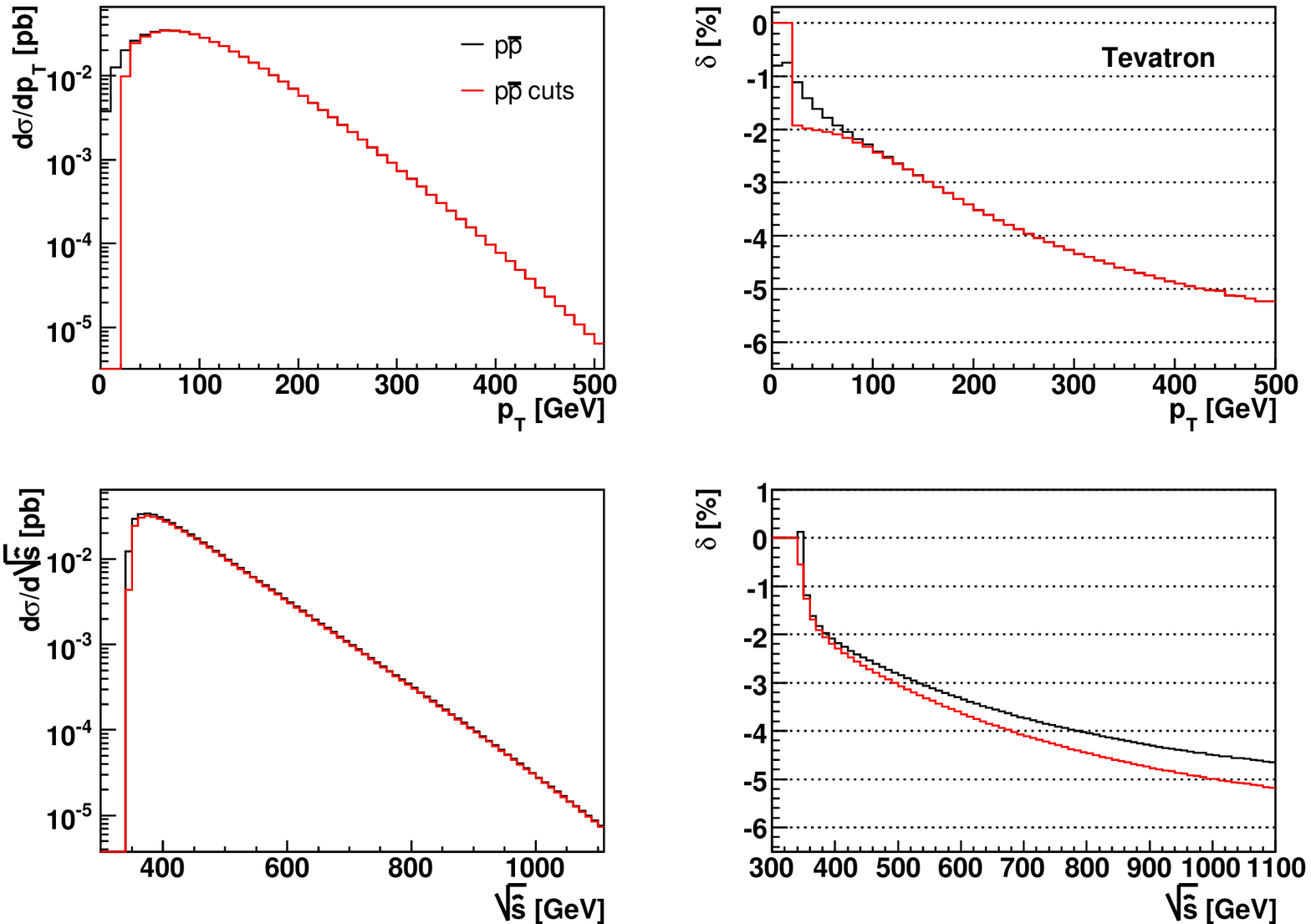}
   \caption{
   Overall NLO QED effects in $\ppbar$ collisions at the Tevatron, for the
   $p_{\rm T}$ distribution (up) and the $\sqrt{\hat s}$ distribution (down),
   without and with application of cuts.
   \label{fig:T_ppbar}}
\end{figure}

Finally, we show the
combination of the partial results for all production subprocesses,
including the photon--gluon channel, in Fig.~\ref{fig:L_pp}
for the LHC and in Fig.~\ref{fig:T_ppbar} for the Tevatron. 
As a consequence of the dominant
$gg$ production channel at the LHC, the total NLO QED corrections 
in the $\sq$ distribution are positive and at the level of about 1\%. 
After applying the cuts and in the $\pt$ distribution the corrections
become negative and tend to increase in size with $\pt$ and $\sq$.
This is caused by the logarithmic final state radiation contribution 
which is not canceled by the virtual corrections in case of 
non-inclusive quantities.

At the Tevatron, with the dominating $\qqbar$ channels,
the total NLO QED corrections have a larger impact owing to the
subtleties of the QED--QCD interference, 
which is not present at the Born-level.
They are negative and in size of several per cent, becoming larger
with increasing $\pt$ and $\sq$, and are slightly enhanced by 
the application of cuts.

\section{Conclusion}
We have provided the last missing item for a complete
EW one-loop calculation for $\ttbar$ production at hadron colliders.
The NLO QED contributions form, together with the non-photonic EW
contributions, the complete EW  
corrections to $\ttbar$ production at the one-loop level.
For consistency and IR-finiteness, 
interference terms between QED and QCD have to be taken
into account, for both virtual and real (bremsstrahlung) contributions.
Moreover, a new class of photon-induced $\ttbar$ production parton processes
occurs, which for the LHC yield larger effects than the
corrections to $\qqbar$ annihilation and $gg$ fusion.
In size, the NLO QED contributions can reach the level of 5\%.
When combined with the rest of the EW corrections, the effects
can become significantly large to find consideration for precision
studies.

\clearpage

\clearpage

\section*{Appendix:  Photon/gluon bremsstrahlung}

In the phase space slicing approach the phase space is divided into a region
where the integrand is finite and into regions where the singularities occur.
In the non-singular case the integration is performed numerically whereas
in the singular regions the integration is carried out analytically 
in the approximation that the photon or photon-like gluon 
is soft and/or collinear to a charged fermion.

We separate the soft and collinear part of the singular regions by 
introducing two cutoff parameters $\Delta E$ and $\Delta\theta$.
In the soft part the photon/gluon energy $k^0$ satisfies the condition
$k^0 < \Delta E \ll \sqrt{\hat{s}}$, while in the collinear
part we have 
$k^0 > \Delta E$ and $\theta_{\gamma f} < \Delta\theta $, where
$\theta_{\gamma f}$ is the angle between the photon and a charged fermion.

In both regions the squared matrix elements for the radiative process
factorize into the lowest-order matrix elements and universal factors
containig the singularities.
Thus, we can decompose the real corrections into
\begin{equation}
\label{slicing}
d\hat{\sigma}^a_{real} = d\hat{\sigma}^a_{soft} + 
d\hat{\sigma}^a_{coll} +  d\hat{\sigma}^a_{finite},
\end{equation}
where $a = q\bar{q}, \, gg$. 
The collinear contribution 
is zero in $gg$ fusion, and in the case of $\qqbar$ annihilation only  
initial state radiation contributes since there are no mass singularities
related to the final state bremsstrahlung.

The soft part is combined with IR and mass singular
virtual corrections to cancel
the IR singularities proportional to $\log \lambda$ and the mass
singularities of double logarithms $\log^2 m_q$. The single
logarithms $\log m_q$ are not compensated in the sum of virtual and
real corrections and have to be handled by means of factorization.
Single virtual and real contributions are dependent on the cutoff 
parameters $\Delta E$ and $\Delta \theta$. However, the dependence has 
to cancel in the combination (\ref{slicing}). 

The soft photon/gluon bremsstrahlung cross sections can be factorized
into soft factors and the Born cross sections,
\begin{eqnarray}   
d\hat{\sigma}^{\qqbar}_{soft} 
& = & -\frac{\alpha}{\pi}\,  d\hat{\sigma}^{\qqbar}_{Born}  \cdot
\left( C_q +C_t +C_{qt} \right) \, ,   \nonumber \\
d\hat{\sigma}^{gg}_{soft} 
& = & -\frac{\alpha}{\pi} \,  d\hat{\sigma}^{gg}_{Born}  \cdot
C_t \, ,   
\end{eqnarray}
with $C_q$, $C_t$, $C_{qt}$ referring to initial state radiation,
final state radiation, and initial--final state radiation interference,
respectively,
\begin{eqnarray}
C_q & = & 
 Q_q^2 \Bigg[  2\loglambda  + 2\loglambda\logmq   
+ \frac{1}{2}\logmqs \nonumber \\ 
&&+ \logmq + \frac{\pi^2}{3} \Bigg] ,    \nonumber   
\end{eqnarray}

\begin{eqnarray}
C_t & = & 
 Q_t^2 \Bigg[ 2\loglambda  + \frac{1}{\beta}\logbeta  
\nonumber \\ &&+  
\frac{\hat s -2m_t^2}{\hat s \beta} \Bigg\{  2\loglambda\logbeta   
\nonumber \\ &&+   
 \frac{1}{2}\logbetas  + 2\li   \Bigg\} \,  \Bigg] \, , \nonumber \\   
C_{qt}&= & 2  Q_q  Q_t  \cdot 3  \Bigg[  2\loglambda\logtu 
\nonumber \\  && + \, \litplus   +  \litmin  \nonumber \\  
&& -\, \liuplus -\liumin \Bigg] ,   
\label{eq:soft}  
\end{eqnarray}
with
$\beta = \sqrt{1-4m_t^2/\hat s}$.
The additional factor of 3
in the interference term comes from the gluon radiation contribution.

The collinear part of
initial state radiation arises only for photons from
the $\qqbar$ annihilation
channel, and can be expressed as follows
(see e.g.~\cite{Dittmaier:2001ay}),
\begin{equation}
d\hat{\sigma}^{\qqbar}_{coll}(\hat{s}) = \frac{Q_q^2\alpha}{\pi}
\int_0^{1-2\Delta E/\sqrt{\hat{s}}} dz \;
d\hat{\sigma}^a_{Born}(z\hat{s})
\left\{ \left[\ln\left(\frac{\Delta\theta^2\hat{s}}{4m_q^2}\right) - 
1 \right] P_{qq}(z) + (1-z) \right\},
\end{equation}
with the splitting function
\begin{equation}
P_{qq}(z)=\frac{1+z^2}{1-z}.
\end{equation}

\clearpage

\bibliographystyle{jhep}
\bibliography{text}

\providecommand{\href}[2]{#2}\begingroup\begin{thebibliography}{10}

\bibitem{Abe:1995hr}
{ CDF} Collaboration, F.~Abe {\em et~al.}, { Observation of top quark
  production in $p\bar{p}$ collisions},  {\em Phys. Rev. Lett.} {\bf 74} (1995)
  2626--2631, [\href{http://xxx.lanl.gov/abs/hep-ex/9503002}{{\tt
  hep-ex/9503002}}].

\bibitem{Abachi:1995iq}
{ D0} Collaboration, S.~Abachi {\em et~al.}, { Observation of the top quark},
  {\em Phys. Rev. Lett.} {\bf 74} (1995) 2632--2637,
  [\href{http://xxx.lanl.gov/abs/hep-ex/9503003}{{\tt hep-ex/9503003}}].

\bibitem{Group:2006mx}
{ The LEP Collaborations ALEPH, DELPHI, L3, OPAL and the LEP Electroweak
  Working Group} Collaboration, { A combination of preliminary electroweak
  measurements and constraints on the standard model},
  \href{http://xxx.lanl.gov/abs/hep-ex/0612034}{{\tt hep-ex/0612034}}.

\bibitem{Gluck:1977zm}
M.~Gl$\ddot{\mathrm u}$ck, J.~F. Owens, and E.~Reya, { Gluon contribution to
  hadronic $J/\psi$ production},  {\em Phys. Rev.} {\bf D17} (1978) 2324.

\bibitem{Combridge:1978kx}
B.~L. Combridge, { Associated production of heavy flavor states in $pp$ and
  $p\bar p$ interactions: some QCD estimates},  {\em Nucl. Phys.} {\bf B151}
  (1979) 429.

\bibitem{Babcock:1977fi}
J.~Babcock, D.~W. Sivers, and S.~Wolfram, { QCD estimates for heavy particle
  production},  {\em Phys. Rev.} {\bf D18} (1978) 162.

\bibitem{Hagiwara:1978hw}
K.~Hagiwara and T.~Yoshino, { Hadroproduction of heavy quark flavors in QCD},
  {\em Phys. Lett.} {\bf B80} (1979) 282.

\bibitem{Jones:1977wu}
L.~M. Jones and J.~Wyld, H.~W., { Production of bound quark - anti-quark
  systems},  {\em Phys. Rev.} {\bf D17} (1978) 2332--2337.

\bibitem{Georgi:1978kx}
H.~M. Georgi, S.~L. Glashow, M.~E. Machacek, and D.~V. Nanopoulos, { Charmed
  particles from two - gluon annihilation in proton proton collisions},  {\em
  Ann. Phys.} {\bf 114} (1978) 273.

\bibitem{Baur:1989qt}
U.~Baur, E.~W.~N. Glover, and A.~D. Martin, { Electroweak interference effects
  in two jet production at $p\bar p$ colliders},  {\em Phys. Lett.} {\bf B232}
  (1989) 519.

\bibitem{Nason:1987xz}
P.~Nason, S.~Dawson, and R.~K. Ellis, { The total cross-section for the
  production of heavy quarks in hadronic collisions},  {\em Nucl. Phys.} {\bf
  B303} (1988) 607.

\bibitem{Nason:1989zy}
P.~Nason, S.~Dawson, and R.~K. Ellis, { The one particle inclusive differential
  cross-section for heavy quark production in hadronic collisions},  {\em Nucl.
  Phys.} {\bf B327} (1989) 49--92.

\bibitem{Beenakker:1988bq}
W.~Beenakker, H.~Kuijf, W.~L. van Neerven, and J.~Smith, { QCD corrections to
  heavy quark production in p anti-p collisions},  {\em Phys. Rev.} {\bf D40}
  (1989) 54--82.

\bibitem{Beenakker:1990ma}
W.~Beenakker, W.~L. van Neerven, R.~Meng, G.~A. Schuler, and J.~Smith, { QCD
  corrections to heavy quark production in hadron hadron collisions},  {\em
  Nucl. Phys.} {\bf B351} (1991) 507--560.

\bibitem{Fadin:1987wz}
V.~S. Fadin and V.~A. Khoze, { Threshold behavior of heavy top production in
  $e^+e^-$ collisions},  {\em JETP Lett.} {\bf 46} (1987) 525--529.

\bibitem{Fadin:1990wx}
V.~S. Fadin, V.~A. Khoze, and T.~Sj$\ddot{\mathrm o}$strand, { On the threshold
  behavior of heavy top production},  {\em Z. Phys.} {\bf C48} (1990) 613--622.

\bibitem{Berger:1996ad}
E.~L. Berger and H.~Contopanagos, { The perturbative resummed series for top
  quark production in hadron reactions},  {\em Phys. Rev.} {\bf D54} (1996)
  3085--3113, [\href{http://xxx.lanl.gov/abs/hep-ph/9603326}{{\tt
  hep-ph/9603326}}].

\bibitem{Berger:1997gz}
E.~L. Berger and H.~Contopanagos, { Threshold resummation of the total cross
  section for heavy quark production in hadronic collisions},  {\em Phys. Rev.}
  {\bf D57} (1998) 253--264,
  [\href{http://xxx.lanl.gov/abs/hep-ph/9706206}{{\tt hep-ph/9706206}}].

\bibitem{Catani:1996dj}
S.~Catani, M.~L. Mangano, P.~Nason, and L.~Trentadue, { The top cross section
  in hadronic collisions},  {\em Phys. Lett.} {\bf B378} (1996) 329--336,
  [\href{http://xxx.lanl.gov/abs/hep-ph/9602208}{{\tt hep-ph/9602208}}].

\bibitem{Catani:1996yz}
S.~Catani, M.~L. Mangano, P.~Nason, and L.~Trentadue, { The resummation of soft
  gluon in hadronic collisions},  {\em Nucl. Phys.} {\bf B478} (1996) 273--310,
  [\href{http://xxx.lanl.gov/abs/hep-ph/9604351}{{\tt hep-ph/9604351}}].

\bibitem{Kidonakis:1995wz}
N.~Kidonakis and J.~Smith, { Top quark inclusive differential distributions},
  {\em Phys. Rev.} {\bf D51} (1995) 6092--6102,
  [\href{http://xxx.lanl.gov/abs/hep-ph/9502341}{{\tt hep-ph/9502341}}].

\bibitem{Laenen:1991af}
E.~Laenen, J.~Smith, and W.~L. van Neerven, { All order resummation of soft
  gluon contributions to heavy quark production in hadron hadron collisions},
  {\em Nucl. Phys.} {\bf B369} (1992) 543--599.

\bibitem{Laenen:1993xr}
E.~Laenen, J.~Smith, and W.~L. van Neerven, { Top quark production
  cross-section},  {\em Phys. Lett.} {\bf B321} (1994) 254--258,
  [\href{http://xxx.lanl.gov/abs/hep-ph/9310233}{{\tt hep-ph/9310233}}].

\bibitem{Bonciani:1998vc}
R.~Bonciani, S.~Catani, M.~L. Mangano, and P.~Nason, { NLL resummation of the
  heavy-quark hadroproduction cross- section},  {\em Nucl. Phys.} {\bf B529}
  (1998) 424--450, [\href{http://xxx.lanl.gov/abs/hep-ph/9801375}{{\tt
  hep-ph/9801375}}].

\bibitem{Cacciari:2003fi}
M.~Cacciari, S.~Frixione, M.~L. Mangano, P.~Nason, and G.~Ridolfi, { The t
  anti-t cross-section at 1.8-TeV and 1.96-TeV: A study of the systematics due
  to parton densities and scale dependence},  {\em JHEP} {\bf 04} (2004) 068,
  [\href{http://xxx.lanl.gov/abs/hep-ph/0303085}{{\tt hep-ph/0303085}}].

\bibitem{Kidonakis:2003qe}
N.~Kidonakis and R.~Vogt, { Next-to-next-to-leading order soft-gluon
  corrections in top quark hadroproduction},  {\em Phys. Rev.} {\bf D68} (2003)
  114014, [\href{http://xxx.lanl.gov/abs/hep-ph/0308222}{{\tt
  hep-ph/0308222}}].

\bibitem{Kidonakis:2005kz}
N.~Kidonakis, { Next-to-next-to-next-to-leading-order soft-gluon corrections in
  hard-scattering processes near threshold},  {\em Phys. Rev.} {\bf D73} (2006)
  034001, [\href{http://xxx.lanl.gov/abs/hep-ph/0509079}{{\tt
  hep-ph/0509079}}].

\bibitem{Beenakker:1993yr}
W.~Beenakker, A.~Denner, W.~Hollik, R.~Mertig, T.~Sack, and D.~Wackeroth, {
  Electroweak one loop contributions to top pair production in hadron
  colliders},  {\em Nucl. Phys.} {\bf B411} (1994) 343--380.

\bibitem{Kao:1997bs}
C.~Kao, G.~A. Ladinsky, and C.~P. Yuan, { Leading weak corrections to the
  production of heavy top quarks at hadron colliders},  {\em Int. J. Mod.
  Phys.} {\bf A12} (1997) 1341--1372.

\bibitem{Kuhn:2005it}
J.~H. K$\ddot{\mathrm u}$hn, A.~Scharf, and P.~Uwer, { Electroweak corrections
  to top-quark pair production in quark-antiquark annihilation},  {\em Eur.
  Phys. J.} {\bf C45} (2006) 139--150,
  [\href{http://xxx.lanl.gov/abs/hep-ph/0508092}{{\tt hep-ph/0508092}}].

\bibitem{Moretti:2006nf}
S.~Moretti, M.~R. Nolten, and D.~A. Ross, { Weak corrections to gluon-induced
  $t\bar t$ hadro-production},  {\em Phys. Lett.} {\bf B639} (2006) 513--519,
  [\href{http://xxx.lanl.gov/abs/hep-ph/0603083}{{\tt hep-ph/0603083}}].

\bibitem{Bernreuther:2006vg}
W.~Bernreuther, M.~Fuecker, and Z.-G. Si, { Weak interaction corrections to
  hadronic top quark pair production},  {\em Phys. Rev.} {\bf D74} (2006)
  113005, [\href{http://xxx.lanl.gov/abs/hep-ph/0610334}{{\tt
  hep-ph/0610334}}].

\bibitem{Kuhn:2006vh}
J.~H. K$\ddot{\mathrm u}$hn, A.~Scharf, and P.~Uwer, { Electroweak effects in
  top-quark pair production at hadron colliders},
  \href{http://xxx.lanl.gov/abs/hep-ph/0610335}{{\tt hep-ph/0610335}}.

\bibitem{Stange:1993td}
A.~Stange and S.~Willenbrock, { Yukawa correction to top quark production at
  the Tevatron},  {\em Phys. Rev.} {\bf D48} (1993) 2054--2061,
  [\href{http://xxx.lanl.gov/abs/hep-ph/9302291}{{\tt hep-ph/9302291}}].

\bibitem{Zhou:1996dx}
H.-Y. Zhou, C.-S. Li, and Y.-P. Kuang, { Yukawa corrections to top quark
  production at the LHC in two-Higgs-doublet models},  {\em Phys. Rev.} {\bf
  D55} (1997) 4412--4420, [\href{http://xxx.lanl.gov/abs/hep-ph/9603435}{{\tt
  hep-ph/9603435}}].

\bibitem{Alam:1996mh}
S.~Alam, K.~Hagiwara, S.~Matsumoto, K.~Hagiwara, and S.~Matsumoto, { One loop
  supersymmetric QCD radiative corrections to the top quark production in p
  anti-p collisions. (Revised version)},  {\em Phys. Rev.} {\bf D55} (1997)
  1307--1315, [\href{http://xxx.lanl.gov/abs/hep-ph/9607466}{{\tt
  hep-ph/9607466}}].

\bibitem{Sullivan:1996ry}
Z.~Sullivan, { Supersymmetric QCD correction to top-quark production at the
  Tevatron},  {\em Phys. Rev.} {\bf D56} (1997) 451--457,
  [\href{http://xxx.lanl.gov/abs/hep-ph/9611302}{{\tt hep-ph/9611302}}].

\bibitem{Li:1996jf}
C.-S. Li, H.-Y. Zhou, Y.-L. Zhu, and J.-M. Yang, { Strong supersymmetric
  quantum effects on top quark production at the Fermilab Tevatron},  {\em
  Phys. Lett.} {\bf B379} (1996) 135--140,
  [\href{http://xxx.lanl.gov/abs/hep-ph/9606271}{{\tt hep-ph/9606271}}].

\bibitem{Li:1995fj}
C.-S. Li, B.-Q. Hu, J.-M. Yang, and C.-G. Hu, { Supersymmetric QCD corrections
  to top quark production in p anti-p collisions},  {\em Phys. Rev.} {\bf D52}
  (1995) 5014--5017.

\bibitem{Kim:1996nz}
J.~Kim, J.~L. Lopez, D.~V. Nanopoulos, and R.~Rangarajan, { Enhanced
  supersymmetric corrections to top-quark production at the Tevatron},  {\em
  Phys. Rev.} {\bf D54} (1996) 4364--4373,
  [\href{http://xxx.lanl.gov/abs/hep-ph/9605419}{{\tt hep-ph/9605419}}].

\bibitem{Zhou:1997fw}
H.-Y. Zhou and C.-S. Li, { Supersymmetric QCD corrections to top quark pair
  production at CERN LHC},  {\em Phys. Rev.} {\bf D55} (1997) 4421--4429.

\bibitem{Berge:2007dz}
S.~Berge, W.~Hollik, W.~M. M$\ddot{\mathrm o}$sle, and D.~Wackeroth, { SUSY QCD
  one-loop effects in (un)polarized top-pair production at hadron colliders},
  {\em to appear in Phys. Rev.}
  [\href{http://xxx.lanl.gov/abs/hep-ph/0703016}{{\tt hep-ph/0703016}}].

\bibitem{Hollik:1997hm}
W.~Hollik, W.~M. M$\ddot{\mathrm o}$sle, and D.~Wackeroth, { Top pair
  production at hadron colliders in non-minimal standard models},  {\em Nucl.
  Phys.} {\bf B516} (1998) 29--54,
  [\href{http://xxx.lanl.gov/abs/hep-ph/9706218}{{\tt hep-ph/9706218}}].

\bibitem{Kao:1999kj}
C.~Kao and D.~Wackeroth, { Parity violating asymmetries in top pair production
  at hadron colliders},  {\em Phys. Rev.} {\bf D61} (2000) 055009,
  [\href{http://xxx.lanl.gov/abs/hep-ph/9902202}{{\tt hep-ph/9902202}}].

\bibitem{Kublbeck:1990xc}
J.~K$\ddot{\mathrm u}$blbeck, M.~B$\ddot{\mathrm o}$hm, and A.~Denner, { Feyn
  Arts: Computer algebraic generation of Feynman graphs and amplitudes},  {\em
  Comput. Phys. Commun.} {\bf 60} (1990) 165--180.

\bibitem{Hahn:2000kx}
T.~Hahn, { Generating Feynman diagrams and amplitudes with FeynArts 3},  {\em
  Comput. Phys. Commun.} {\bf 140} (2001) 418--431,
  [\href{http://xxx.lanl.gov/abs/hep-ph/0012260}{{\tt hep-ph/0012260}}].

\bibitem{Hahn:2001rv}
T.~Hahn and C.~Schappacher, { The implementation of the minimal supersymmetric
  standard model in FeynArts and FormCalc},  {\em Comput. Phys. Commun.} {\bf
  143} (2002) 54--68, [\href{http://xxx.lanl.gov/abs/hep-ph/0105349}{{\tt
  hep-ph/0105349}}].

\bibitem{Hahn:1998yk}
T.~Hahn and M.~Perez-Victoria, { Automatized one-loop calculations in four and
  D dimensions},  {\em Comput. Phys. Commun.} {\bf 118} (1999) 153--165,
  [\href{http://xxx.lanl.gov/abs/hep-ph/9807565}{{\tt hep-ph/9807565}}].

\bibitem{Hahn:1999mt}
T.~Hahn, { Loop calculations with FeynArts, FormCalc, and LoopTools},  {\em
  Acta Phys. Polon.} {\bf B30} (1999) 3469--3475,
  [\href{http://xxx.lanl.gov/abs/hep-ph/9910227}{{\tt hep-ph/9910227}}].

\bibitem{Hahn:2000jm}
T.~Hahn, { Automatic loop calculations with FeynArts, FormCalc, and LoopTools},
   {\em Nucl. Phys. Proc. Suppl.} {\bf 89} (2000) 231--236,
  [\href{http://xxx.lanl.gov/abs/hep-ph/0005029}{{\tt hep-ph/0005029}}].

\bibitem{Hahn:2006qw}
T.~Hahn and M.~Rauch, { News from FormCalc and LoopTools},  {\em Nucl. Phys.
  Proc. Suppl.} {\bf 157} (2006) 236--240,
  [\href{http://xxx.lanl.gov/abs/hep-ph/0601248}{{\tt hep-ph/0601248}}].

\bibitem{'tHooft:1978xw}
G.~'t~Hooft and M.~J.~G. Veltman, { Scalar one loop integrals},  {\em Nucl.
  Phys.} {\bf B153} (1979) 365--401.

\bibitem{Passarino:1978jh}
G.~Passarino and M.~J.~G. Veltman, { One loop corrections for $e^+ e^-$
  annihilation into $\mu^+\mu^-$ in the Weinberg model},  {\em Nucl. Phys.}
  {\bf B160} (1979) 151.

\bibitem{Beenakker:1988jr}
W.~Beenakker and A.~Denner, { Infrared divergent scalar box integrals with
  applications in the electroweak standard model},  {\em Nucl. Phys.} {\bf
  B338} (1990) 349--370.

\bibitem{Denner:1991qq}
A.~Denner, U.~Nierste, and R.~Scharf, { A compact expression for the scalar one
  loop four point function},  {\em Nucl. Phys.} {\bf B367} (1991) 637--656.

\bibitem{Beenakker:1991ca}
W.~Beenakker, S.~C. van~der Marck, and W.~Hollik, { $e^+e^-$ annihilation into
  heavy fermion pairs at high-energy colliders},  {\em Nucl. Phys.} {\bf B365}
  (1991) 24--78.

\bibitem{Bloch:1937pw}
F.~Bloch and A.~Nordsieck, { Note on the radiation field of the electron},
  {\em Phys. Rev.} {\bf 52} (1937) 54--59.

\bibitem{Dittmaier:1999mb}
S.~Dittmaier, { A general approach to photon radiation off fermions},  {\em
  Nucl. Phys.} {\bf B565} (2000) 69--122,
  [\href{http://xxx.lanl.gov/abs/hep-ph/9904440}{{\tt hep-ph/9904440}}].

\bibitem{Catani:1996jh}
S.~Catani and M.~H. Seymour, { The dipole formalism for the calculation of QCD
  jet cross sections at next-to-leading order},  {\em Phys. Lett.} {\bf B378}
  (1996) 287--301, [\href{http://xxx.lanl.gov/abs/hep-ph/9602277}{{\tt
  hep-ph/9602277}}].

\bibitem{Catani:1996vz}
S.~Catani and M.~H. Seymour, { A general algorithm for calculating jet cross
  sections in NLO QCD},  {\em Nucl. Phys.} {\bf B485} (1997) 291--419,
  [\href{http://xxx.lanl.gov/abs/hep-ph/9605323}{{\tt hep-ph/9605323}}].

\bibitem{Martin:2004dh}
A.~D. Martin, R.~G. Roberts, W.~J. Stirling, and R.~S. Thorne, { Parton
  distributions incorporating QED contributions},  {\em Eur. Phys. J.} {\bf
  C39} (2005) 155--161, [\href{http://xxx.lanl.gov/abs/hep-ph/0411040}{{\tt
  hep-ph/0411040}}].

\bibitem{Baur:1998kt}
U.~Baur, S.~Keller, and D.~Wackeroth, { Electroweak radiative corrections to W
  boson production in hadronic collisions},  {\em Phys. Rev.} {\bf D59} (1999)
  013002, [\href{http://xxx.lanl.gov/abs/hep-ph/9807417}{{\tt
  hep-ph/9807417}}].

\bibitem{Diener:2005me}
K.~P.~O. Diener, S.~Dittmaier, and W.~Hollik, { Electroweak higher-order
  effects and theoretical uncertainties in deep-inelastic neutrino scattering},
   {\em Phys. Rev.} {\bf D72} (2005) 093002,
  [\href{http://xxx.lanl.gov/abs/hep-ph/0509084}{{\tt hep-ph/0509084}}].

\bibitem{Dittmaier:2001ay}
S.~Dittmaier and M.~Kr$\ddot{\mathrm a}$mer, { Electroweak radiative
  corrections to W-boson production at hadron colliders},  {\em Phys. Rev.}
  {\bf D65} (2002) 073007, [\href{http://xxx.lanl.gov/abs/hep-ph/0109062}{{\tt
  hep-ph/0109062}}].

\end{thebibliography}\endgroup

\end{document}